\documentclass[aps,prb,twocolumn,a4paper,floatfix,showpacs,superscriptaddress]{revtex4}
\usepackage{graphicx}
\usepackage{hyperref}
\usepackage{amsmath}
\usepackage{lineno}
\usepackage{verbatim}

\usepackage[normalem]{ulem}

\begin{document}

\title{Comparative study of first-principles approaches for effective Coulomb interaction strength $U_{\rm eff}$ between localized $f$-electrons: lanthanide metals as an example}

\author{Bei-Lei Liu}
\affiliation{School of Mathematical Sciences, Beijing Normal University, Beijing 100875, China}
\affiliation{Laboratory of Computational Physics, Institute of Applied Physics and Computational Mathematics, Beijing 100088, China}

\author{Yue-Chao Wang\footnote{Corresponding authors: yuechao\_wang@126.com}}
\affiliation{Laboratory of Computational Physics, Institute of Applied Physics and Computational Mathematics, Beijing 100088, China}

\author{Yu Liu\footnote{Corresponding authors: liu\_yu@iapcm.ac.cn}}
\affiliation{Laboratory of Computational Physics, Institute of Applied Physics and Computational Mathematics, Beijing 100088, China}

\author{Yuan-Ji Xu}
\affiliation{Institute for Applied Physics, University of Science and Technology Beijing, Beijing 100083, China}

\author{Xin Chen}
\affiliation{Laboratory of Computational Physics, Institute of Applied Physics and Computational Mathematics, Beijing 100088, China}

\author{Hong-Zhou Song}
\affiliation{Laboratory of Computational Physics, Institute of Applied Physics and Computational Mathematics, Beijing 100088, China}

\author{Yan Bi}
\affiliation{Center for High Pressure Science and Technology Advanced Research, Beijing 100094, China}
\affiliation{National Key Laboratory for Shock Wave and Detonation Physics, Institute of Fluid Physics,
CAEP, Mianyang 621900, China}

\author{Hai-Feng Liu}
\affiliation{Laboratory of Computational Physics, Institute of Applied Physics and Computational Mathematics, Beijing 100088, China}

\author{Hai-Feng Song}
\affiliation{Laboratory of Computational Physics, Institute of Applied Physics and Computational Mathematics, Beijing 100088, China}

\begin{abstract}

As correlation strength has a key influence on the simulation of strongly correlated materials, many approaches have been proposed to obtain the parameter using first-principles calculations. However, a comparison of the different Coulomb strengths obtained using these approaches and an investigation of the mechanisms behind them are still needed.
Taking lanthanide metals as an example, we research the factors that affect the effective Coulomb interaction strength, $U_{\rm eff}$, by local screened Coulomb correction (LSCC), linear response (LR), and constrained random-phase approximation (cRPA) in VASP.
The $U^{\rm LSCC}_{\rm eff}$ value increases from 4.75 eV to 7.78 eV, $U^{\rm LR}_{\rm eff}$ is almost stable at about 6.0 eV (except for Eu, Er, and Yb), and $U^{\rm cRPA}_{\rm eff}$ shows a two-stage decreasing trend in both light and heavy lanthanides.
To investigate these differences, we establish a scheme to analyze the coexistence and competition between the orbital localization and the screening effect. We find that LSCC and cRPA are dominated by the orbital localization and the screening effect, respectively, whereas LR shows a balance of the competition between the two factors.
Additionally, the performance of these approaches is influenced by different starting points from PBE and PBE+$U$, especially for cRPA. Our results provide useful knowledge for understanding the $U_{\rm eff}$ of lanthanide materials, and similar analyses can also be used in the research of other correlation strength simulation approaches.

\end{abstract}
\maketitle


\section{Introduction}

In recent decades, the class of elements with $3d$, $4f$, or $5f$ electrons and their compounds has attracted tremendous attention in both fundamental research and industrial applications.\cite{Gschneidner1978,Grasso2013,Croat2013} These systems exhibit many exotic properties, such as high-temperature superconductivity\cite{Kamihara2008} and colossal magneto-resistive effect.\cite{Tokura2006} Such materials are called strongly correlated materials due to their strong electron–electron interaction.\cite{Anisimov2010,Avella2012} However, the widely used density functional theory (DFT)\cite{Martin04} in conventional local density approximation (LDA)\cite{Ceperley1980,Perdew1981} or generalized gradient approximation (GGA)\cite{Perdew1996} is not suitable for these materials because of the spurious self-interaction.\cite{Hamada1995,Sarma1995,Amadon2008,Savrasov2000} Many $ab~initio$ methods developed for strongly correlated materials have been successfully used in simulations for thermodynamic properties and electronic structure properties.\cite{Perdew1981,Casadei2016,Perdew1996,Anisimov2010} Among them, correction methods based on the onsite Hubbard model, such as DFT+$U$,\cite{Anisimov1991,Anisimov1997,Himmetoglu2014} DFT plus dynamic mean-field theory
(DFT+DMFT),\cite{Kotliar2006,McMahan2009} and DFT plus Gutzwiller projected variational wave function (DFT+Gutzwiller)\cite{Deng2008,Lanata2015} are popular and effective.
Despite their good performance, these methods still have many key problems in the construction of a high-precision Hubbard model, which mainly include the following: (1) the definition of correlation space, namely, the choice of local projection lacks clarity;\cite{Shick1999,Amadon2008} (2) the double counting term, which needs to be deducted from the DFT part, has no strict form;\cite{Czyzyk1994,Ryee2018,Wang2016} and, especially, (3) the correlation strength between localized electrons, namely, the Hubbard $U$ that describes the on-site Coulomb interaction under a certain screening, and Hund exchange interaction parameter $J$ is needed.\cite{Wang2006,Jiang2012} The correlation strength can even lead to qualitative differences in material properties.\cite{Eryigit2022} In the isotropic form,\cite{Dudarev1998} the correlation strength is directly expressed by the effective Coulomb interaction between localized electrons, $U_{\rm eff} \equiv U-J$.

In experiments, the $U$ value is typically estimated via the positions of Hubbard bands in X-ray photoemission spectroscopy (XPS) and bremsstrahlung isochromat spectroscopy (BIS) data.\cite{Lang1981,Kaurila1997} However, these experiments are costly and difficult to perform under some extreme conditions. In early simulation work, the $U_{\rm eff}$ values were tuned by fitting the simulated properties to experimental ones, such as the band gap.\cite{Bengone2000,Castleton2007,Pegg2017} However, this approach is not applicable in cases where experimental data are lacking. In addition, it reduces the predictive ability of the correction methods based on the Hubbard model, and there may also be significant differences in $U_{\rm eff}$ values as a result of using different reference properties for the same system.

To address the above-mentioned problems, many approaches based on first-principles have been developed to simulate the strength of on-site Coulomb interaction. For example,
(1) Herbst \emph{et al.} evaluated Hubbard-$U$ using Hartree–Fock calculations. However, this method is based on the atomic limit, which does not reflect the systems we really care about.\cite{Qiu2020}
(2) Dederichs \emph{et al.}\cite{Dederichs1984} developed the constrained density functional theory (cDFT) approach to calculate the energy difference by constructing electronic configurations. In this approach, localized electrons are treated as core electrons in the full-potential framework.
(3) Later, Cococcioni and Gironcoli \cite{Cococcioni2005} extended the cDFT approach to the more efficient pseudopotential framework based on linear response (LR) theory, and obtained the $U_{\rm eff}$ value through the perturbation-induced variation of the occupation number of localized electrons.
4) Aryasetiawan \emph{et al.}\cite{Aryasetiawan2004} developed the constrained random-phase approximation (cRPA) method based on many-body perturbation theory, which can calculate frequency-dependent $U$ values. cRPA is generally considered a calculation method that comprehensively involves physical effects,\cite{Tesch2022} but the computational cost is relatively high, and there is uncertainty in many factors, such as the screening channel.
5) Wang and Jiang\cite{Wang2019} proposed the local screened Coulomb correction (LSCC) method in recent years, which has been used to simulate the Coulomb interaction of $5f$-electrons in $\beta$-UH$_{\rm 3}$\cite{Wu2022} and systematically calculate the $U$ value of $3d$ metal oxides.\cite{Wang2019} The method has good performance in evaluating Hubbard-$U$ with dramatically reduced computational cost.

The above approaches are based on different physical processes, and it has been noted that they could yield different $U_{\rm eff}$ values for the same system.\cite{Qiu2020,Tesch2022}
Nawa \emph{et al.}\cite{Nawa2018} noted that comparing the absolute values of $U_{\rm eff}$ calculated by different methods may not be meaningful. However, comparing the characteristics of different methods is important from the perspectives of both theory and practice. Specifically, the changing trend of $U_{\rm eff}$ with atomic number can help us understand such characteristics. For example, Tesch and Kowalski\cite{Tesch2022} researched the changing trend of $U_{\rm eff}$ calculated by the LR method for $d$-electron metals and emphasized the rise of correlation strength caused by stronger localization of $d$ orbitals. Qiu \emph{et al.}\cite{Qiu2020} studied the trend of $U_{\rm eff}$ calculated by the LR method for actinide metals and emphasized the screening effect caused by different structures and the electronic localization change, which was related to the evolution of atomic volume.

Lanthanide metals are typical strongly correlated series; their $4f$ electrons show stronger localization than those of $3d$ and $5f$ orbitals, which creates a good platform for exploring the characteristics of different simulation methods. The correction methods based on the Hubbard model have been widely used to simulate the properties of lanthanide metals. To name a few, Harmon \emph{et al.}\cite{Harmon1995} simulated Gd using LDA+$U$ with $U_{\rm eff} = 6.0$ eV and obtained a correct description of the antiferromagnetic ground state. Mohanta \emph{et al.}\cite{Mohanta2010} simulated Gd-Lu using GGA+$U$ and showed that $U_{\rm eff} = 6.8$ eV could obtain structural and magnetic properties that were basically consistent with the measured data. Locht \emph{et al.}\cite{Locht2016} carried out DFT+DMFT simulation with $U = 7$ eV, and the structural properties, magnetic properties, and photoelectron spectra were in good agreement with the experimental results.
We note that there are also many works using first-principles approaches to calculate the correlation strength of $4f$ electrons. Nilsson \emph{et al.}\cite{Nilsson2013} reported the calculation of the cRPA-$U$ value for Ce-Gd, and Morée and Amadon\cite{Moree2018} carried out a self-consistent calculation by combining LSDA+$U$ with cRPA for lanthanide metals. Results that have been obtained by using the cDFT/LR method to calculate the $U_{\rm eff}$ value of lanthanide metals are scattered across different works. For example, Dederichs \emph{et al.} calculated the $U_{\rm eff}$ of Ce as an impurity atom using cDFT; Cococcioni \emph{et al.}\cite{Cococcioni2005} calculated the $U_{\rm eff}$ of Ce using the LR method; and Tao \emph{et al.}\cite{Tao2014} calculated the $U_{\rm eff}$ value of the Gd$_{13}$ cluster. However, as far as we know, a systematic calculation of $U_{\rm eff}$ values using the cDFT/LR approach has not yet been performed.
The performance of LSCC in lanthanide metals will also be reported for the first time in this paper.

In this work, we systematically simulate the on-site Coulomb interaction between $4f$ electrons of lanthanide metals using the LSCC, LR, and cRPA approaches, all of which are performed in the popular Vienna Ab~initio Simulation Package (VASP).\cite{Kresse1993,Kresse1996}
The changing trends of $U_{\rm eff}$ calculated by different approaches are compared. In particular, we analyze the competition between the orbital localization and the screening effect to explain the different performance of $U_{\rm eff}$. In addition, we detect the sensitivity of $U_{\rm eff}$ to initial electronic state by simulation. 

The paper is organized as follows. In Sec. II, we introduce the methods and the parameter setting used in this work. In Sec. III, the results of effective Coulomb interaction values are given, analyses of orbital localization and screening effect are performed, and the performance of $U_{\rm eff}$ obtained from different electronic states is compared. Finally, our conclusions are summarized in Sec. IV.


\section{Methodology}

\subsection{Local screened Coulomb correction (LSCC) approach}
The LSCC approach calculates the on-site interaction between localized electrons through the screened Coulomb interaction in the form of Yukawa potential:\cite{Wang2019}
\begin{equation}\label{lscc-sc}
v_{sr}({\bf r},{\bf r'}) = \frac{e^{-\lambda|\bf r - r'|}}{|{\bf r - r'}|},
\end{equation}
where $\lambda = 2 [\frac{3\rho}{\pi}]^{1/6}$ is the parameter characterizing screening strength in the Thomas–Fermi screening model. For real systems, we adopt a density-weighted averaging scheme (see S-III type in Wang and Jiang):\cite{Wang2019}
\begin{equation}
\label{TF-lam}
\bar{\lambda} = \frac{\int_{\rm aug} \lambda({\bf r})\rho({\bf r})d{\bf r}}{\int_{\rm aug}\rho({\bf r})d{\bf r}},
\end{equation}
where $\rho({\bf r})$ is the electron-density, and $\lambda({\bf r})$ is the local screening parameter. The result is obtained by weighted-averaging $\lambda({\bf r})$ within the augmentation region.
Using \eqref{lscc-sc}, the screened Slater integrals can be calculated by\cite{Bultmark2009}
\begin{equation}
U_{m_1,m_2,m_3,m_4} = \langle \varphi_{l,m_1} \varphi_{l,m_2} | v_{sr} | \varphi_{l,m_3} \varphi_{l,m_4} \rangle,
\end{equation}
where $|\varphi_{l,m} \rangle$ is the local orbital with angular momentum quantum number $l$ and magnetic quantum number $m$. We construct the radial part of the local orbital as\cite{Novak2006,Wang2019}
\begin{equation}
\label{radi-part}
\phi_l(r) = \sqrt{\frac{\rho_{l0}(r)}{N_l}},
\end{equation}
where $\rho_{l0}(r)$ is the spherically symmetric part of $l$-electron density within the PAW augmentation sphere, and $N_l$ is the occupation number of the corresponding orbitals.
Then, we deduce parameters $U$ and $J$ as follows:
\begin{align}
\label{eff_U}
U &= \frac{1}{(2l+1)^2}
\sum_{m_2=1}^{2l+1}\sum_{m_1=1}^{2l+1} U_{m_1,m_2,m_1,m_2},
\\
\nonumber
J &= U - \frac{1}{(2l+1)(2l)}\sum_{m_2=1}^{2l+1},
\\
\label{eff_J}
&\sum_{m_1=1}^{2l+1} \left[ U_{m_1,m_2,m_1,m_2} - U_{m_1,m_2,m_2,m_1} \right].
\end{align}

\subsection{Linear response (LR) approach}
In the LR approach, the interaction parameter can be calculated as the second derivative of the ground-state total energy with respect to the local electron occupation number, $n^I$, in site $I$.\cite{Cococcioni2005}
The effective electronic potential is perturbed by an external potential $\alpha^{I} |\varphi^{I}_m \rangle \langle \varphi^{I}_m|$ that only acts on the localized orbitals with amplitude $\alpha^{I}$. The second derivative of the total energy with respect to $n^I$ equals the inverse of the response function:
\begin{equation}
\chi_{{I},{J}} = \frac{\delta n^{I}}{\delta \alpha^{J}}.
\end{equation}
In actual calculation, the Kohn-Sham wave functions of non-interacting electron systems would be reorganized in response to the perturbation. This artificial response is not related to the electron interaction and therefore must be subtracted. The response function of this part, $
\chi_0$, could be evaluated by performing a non-self-consistent DFT calculation while keeping the charge density constant.\cite{Cococcioni2005,Qiu2020} The on-site interaction is given by
\begin{equation}
U^{I}_{\rm eff} = (\chi^{-1}_0 - \chi^{-1})_{{I},{I}}.
\end{equation}

\subsection{Constrained random-phase approximation (cRPA) approach}

The cRPA approach assumes that the screening between localized electrons can be well treated in the effective low-energy model, and only the screening between the rest electrons needs to be considered in the calculation of Hubbard $U$.\cite{Aryasetiawan2004, Aryasetiawan1998, Aryasetiawan2006} The effective interaction in the reduced space is defined as
\begin{equation}
W_r({\bf r},{\bf r'};\omega) = \frac{v({\bf r},{\bf r'})}{1 - v({\bf r},{\bf r'})P_r({\bf r},{\bf r'};\omega)},
\end{equation}
where $P_r$ is the polarization (excluding transition processes between correlated states), and $v$ is the bare Coulomb interaction.
From $W_r$, we can calculate the effective interaction matrix as\cite{Moree2018}
\begin{equation}
U_{m_1,m_2,m_3,m_4}(\omega) = \langle \varphi_{l,m_1} \varphi_{l,m_2} | W_r(\omega) | \varphi_{l,m_3} \varphi_{l,m_4} \rangle,
\end{equation}
and derive parameters $U$ and $J$ as in Eqs. \eqref{eff_U} and \eqref{eff_J} at $\omega = 0$.

In the random-phase approximation, the full polarization, $P$, is calculated as
\begin{align}
\nonumber
P({\bf r},{\bf r'};\omega) &= \sum_{{\bf k}n}^{occ}\sum_{{\bf k'}n'}^{unocc} \frac{\psi^*_{{\bf k'}n'}({\bf r})\psi_{{\bf k}n}({\bf r})\psi^*_{{\bf k}n}({\bf r'})\psi_{{\bf k'}n'}({\bf r'})}{\omega - \varepsilon_{{\bf k'}n'} + \varepsilon_{{\bf k}n} + i\eta}
\\
\label{screend-polar}
&- \frac{\psi_{{\bf k}n}({\bf r})\psi^*_{{\bf k'}n'}({\bf r})\psi_{{\bf k'}n'}({\bf r'})\psi^*_{{\bf k}n}({\bf r'})}{\omega + \varepsilon_{{\bf k'}n'} - \varepsilon_{{\bf k}n} - i\eta},
\end{align}
where $\eta$ is a positive infinitesimal.
The polarizability between localized $f$ electrons, $P_f$, is calculated for the disentangled band structure according to Eq. \eqref{screend-polar}, and the polarizability of the rest subspace is then calculated by $P_r = P - P_f$. If the correlated bands are completely isolated from the other bands, $P_f$ can be straightforwardly calculated according to band indices. For the case where the correlated bands are entangled, there are currently some disentanglement schemes that can be used.\cite{Miyake2009,Sasioglu2011,Kaltak2015} We adopt the disentangling method\cite{Miyake2009} in our implementation. In this scheme, all coupling between the correlated and rest subspaces in the Hamiltonian are removed, and then the Hamiltonian is diagonalized in these subspaces. The polarizabilities $P$ and $P_f$ are then calculated by the disentangled wavefunctions. Recent works\cite{Amadon2014,Moree2018} have shown that this scheme can better calculate $U$ values from the DFT band structure compared to the weighted method. However, the disentanglement method has its limitation, the dependence of the energy window.

\subsection{Computational details}

The effective Coulomb interaction strengths, $U_{\rm eff}$, are calculated for a lanthanide metal (Ce-Yb) with fcc structure, whose lattice parameters are set as the experimental values at room temperature (see the Appendix). All calculations have been performed using VASP.\cite{Kresse1993,Kresse1996} Projector augmented wave (PAW) pseudopotentials\cite{Blochl1994} are used. Please see the Appendix for details about the PAW parameters. The $f$ electrons are treated as valence electrons, and the Perdew–Burke–Ernzerhof (PBE)\cite{Perdew1996} functional is applied for all calculations. We calculate $U_{\rm eff}$ values for all systems in paramagnetic states. All the plane-wave energy cutoffs are set as 600 eV. A Monkhorst-Pack\cite{Monkhorst1976} k-mesh of $8 \times 8 \times 8$ (the k-mesh in the cRPA calculation is set as $4 \times 4 \times 4$, which meets a precision of 0.2 eV for $U_{\rm eff}$ value) is applied to ensure convergence. The LR approach is performed on $2 \times 2 \times 2$ supercells. Limited by the current implementation in the VASP code, the same local orbitals in LSCC and LR cannot be chosen in cRPA. We set the local orbitals as maximally localized Wannier functions\cite{Marzari1997,Souza2001,Mostofi2008} (MLWFs) constructed from a large energy window [$- 20$ eV, $20$ eV], and we use 100 bands in the calculation of the polarizability.


\section{Results and Discussion}
\subsection{Trends of effective Coulomb interaction with different approaches}

\begin{figure}[htbp]
\centering
\includegraphics[width=0.44\textwidth]{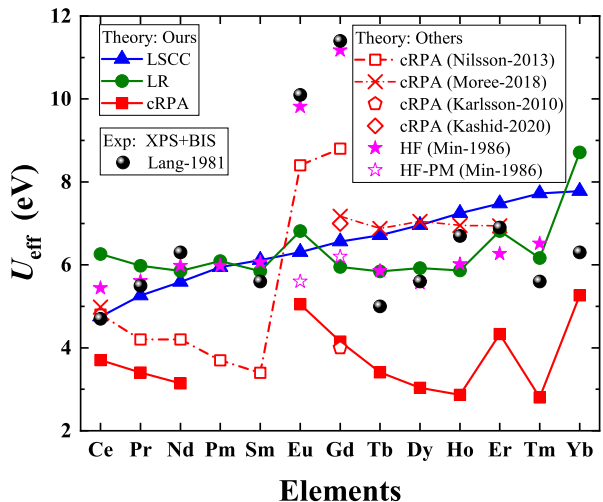}
\caption{The effective Coulomb interaction strength, $U_{\rm eff}$, for lanthanide metals. The blue filled triangles represent the results obtained with the LSCC approach, the green filled circles represent the results obtained with the LR approach, and the red symbols represent the results obtained with the cRPA approach (filled squares for this work, open squares for results from Nilsson \emph{et al.},\cite{Nilsson2013} crosses for results from Morée and Amadon,\cite{Moree2018} pentagons for results from Karlsson \emph{et al.},\cite{Karlsson2010} and diamonds for results from Kashid\emph{et al.}\cite{Kashid2020}). The magenta symbols represent the results obtained with Hartree–Fock (HF) calculation (open stars for ferromagnetic states and filled stars for paramagnetic (PM) states by Min \emph{et al.},\cite{Min1986} and black filled balls for XPS and BIS experiments by Lang \emph{et al.}\cite{Lang1981})}\label{fig:ueff}
\end{figure}

We present the effective Coulomb interaction, $U_{\rm eff}$, derived by the different first-principles approaches and other theoretical and experimental reference data in Fig. \ref{fig:ueff}. In the following, we denote the $U_{\rm eff}$ calculated by the LSCC (or LR, cRPA) approach as $U^{\rm LSCC}_{\rm eff}$ (or $U^{\rm LR}_{\rm eff}$, $U^{\rm cRPA}_{\rm eff}$). Our calculations show that the $U_{\rm eff}$ values obtained by the different approaches have different changing trends with increasing atomic number. Specifically, $U^{\rm LSCC}_{\rm eff}$ increases from 4.75 eV to 7.78 eV with increasing atomic number.
The $U^{\rm LR}_{\rm eff}$ value is about 6.0 eV, with a maximum variation of 0.26 eV except Eu, Er, and Yb. The $U^{\rm LR}_{\rm eff}$ value for Eu and Er is about 6.8 eV, and that for Yb is significantly larger than the others. The $U^{\rm cRPA}_{\rm eff}$ in this work shows two decreasing trends in light (Ce-Nd) and heavy (Eu-Ho) lanthanides, and also abruptly jumps in Eu, Er, and Yb. $U^{\rm cRPA}_{\rm eff}$ is smaller than $U^{\rm LSCC}_{\rm eff}$ and $U^{\rm LR}_{\rm eff}$ in lanthanides. Some convergence problems occur when we try to use an appropriate number of virtual states for the cRPA calculation. However, results for these two elements are not vital for our conclusion.

The values of $U^{\rm LSCC}_{\rm eff}$ and $U^{\rm LR}_{\rm eff}$ are in the range of previous empirical $U$ values, and, except for Eu and Gd, are closer to the experimental estimates and HF atomic calculation than $U^{\rm cRPA}_{\rm eff}$. We note that this difference may relate to the spin polarization. There are reports of calculation results in the literature obtained while ignoring the spin polarization effect that are close to our results. Moreover, we note that the values with spin polarization are very large (about 10–12 eV), which may not be in the range that can simulate properties that are in good agreement with experiments. For $U^{\rm cRPA}_{\rm eff}$, the decreasing trend in light lanthanides also could be observed in the work of Nilsson \emph{et al.}\cite{Nilsson2013} However, our results for Eu and Gd are quite different from those of Nilsson \emph{et al.} \cite{Nilsson2013} We note that the effect of spin polarization was also considered in Nilsson \emph{et al.}\cite{Nilsson2013}, and the Gd-$U_{\rm eff}$ values reported in the literature show large differences due to the use of different models and numerical methods in cRPA\cite{Karlsson2010,Kashid2020}. Finally, we point out that the $U_{\rm eff}$ value in the work of Morée and Amadon\cite{Moree2018} calculated by a self-consistent cRPA scheme combined with DFT+$U$ is generally larger than ours and does not show an obvious decreasing trend. We will discuss this phenomenon in Section \ref{Sensitive}.

Overall, similar performance of these approaches can be observed in both this work and many other reports. On the one hand, these results confirm that the approaches in this work are correctly used. On the other hand, the trends of $U_{\rm eff}$ displayed in our results is general which may be encountered in other situations. Based on the simulations above, analyses of the different characteristics of $U^{\rm LSCC}_{\rm eff}$, $U^{\rm LR}_{\rm eff}$, and $U^{\rm cRPA}_{\rm eff}$ are given in the next section.

\subsection{Characterization for localization of orbital and screening strength}

To understand the trends of $U_{\rm eff}$, we investigate the evolution of orbital localization and screening effect. From its definition, $U_{\rm eff}$ represents the on-site interaction between local electrons that are screened by the environment. Thus, the orbital localization of $f$-electrons and the screening strength of the environment are two key factors that should influence $U_{\rm eff}$, and a quantitative analysis of these two factors is of great importance to understanding the performance of different simulation approaches.

\begin{figure}
\centering
\includegraphics[width=0.49\textwidth]{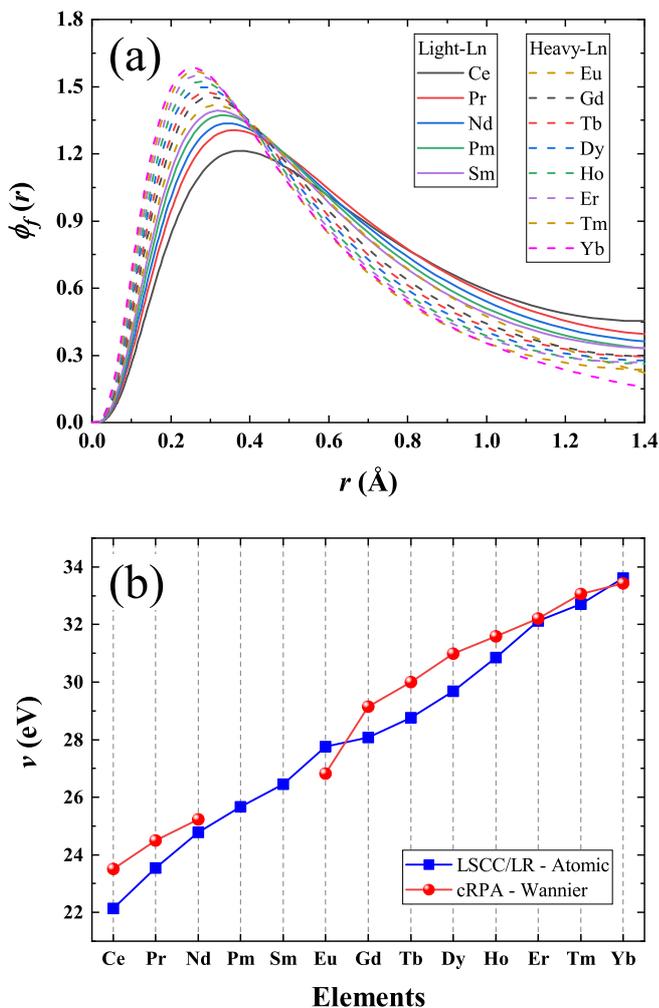}
\caption{(a) Radial $4f$ orbitals $\phi_{f}(r)$ of lanthanides in LSCC and LR approaches, where the solid lines represent light lanthanides, and the dashed lines represent heavy lanthanides. (b) Bare Coulomb interaction, $v$, of $4f$ orbitals in (a) and Wannier functions in the cRPA approach, represented as blue squares and red circles, respectively.}\label{fig:local}
\end{figure}

First, the orbital localization of $f$-electrons is shown in Fig. \ref{fig:local}. In Fig. \ref{fig:local}(a), the radial part of the $4f$-orbitals used in the LSCC and LR approaches is shown. With the increase of atomic number, the $f$-electrons in real space become more localized. To quantitatively characterize the orbital localization, the bare Coulomb interaction, $v$, is used, because localized electron distribution directly relates to a large $v$. The $v$ of LSCC/LR and cRPA are shown in Fig. \ref{fig:local}(b). As the atomic number increases, $v$ increases from 22.13 eV (23.51 eV for cRPA) to 35.40 eV (35.14 eV for cRPA). We note that bare Coulomb interactions of these two kinds of orbitals have a difference of less than 6\%, which indicates that the selected energy window gives a relatively good coherence between the atomic $4f$-orbitals and Wannier functions. To conclude, the results above show that with the increase of atomic number, the localization of $4f$-orbitals becomes stronger, which will lead to an increased $U_{\rm eff}$ value.

\begin{figure}
\centering
\includegraphics[width=0.44\textwidth]{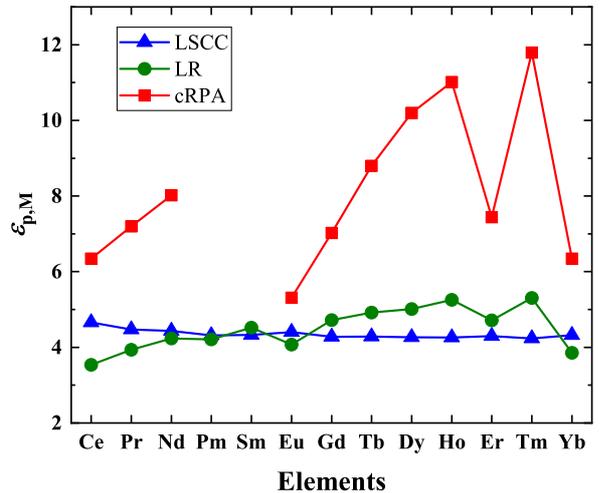}
\caption{The macroscopic dielectric constant $\varepsilon_{\rm p,M} = \frac{v}{U_{\rm eff}}$ of lanthanides. The blue triangles represent the results of the LSCC approach, the green circles represent the results of the LR approach, and the red squares represent the results of the cRPA approach.}\label{fig:diele}
\end{figure}

Second, the screening strength is characterized in this work by the macroscopic dielectric constant $\varepsilon_{\rm p,M} = \frac{v}{U_{\rm eff}}$. It should be mentioned that $\varepsilon_{\rm p,M}$ differs from the real dielectric constant because $U_{\rm eff}$ is the average interaction of on-site electrons, not the interaction at the static limit. However, this definition can intuitively reflect the screening effect from the different $U_{\rm eff}$ simulation approaches. As seen in Fig. \ref{fig:diele}, the results of LSCC change little within the $4f$ elements, with a slight decrease of 0.5. $\varepsilon^{\rm LR}_{\rm p,M}$ shows an enhanced trend—except for Eu, Er, and Yb—with a variation of less than 1.8, whereas $\varepsilon^{\rm cRPA}_{\rm p,M}$ is the largest in absolute value and amplitude change, showing a two-stage enhanced trend in light and heavy lanthanides. Both $\varepsilon^{\rm LR}_{\rm p,M}$ and $\varepsilon^{\rm cRPA}_{\rm p,M}$ decrease suddenly in Eu, Er, and Yb. These results show that the increasing trend of $U_{\rm eff}$ value in LSCC is dominated by the orbital localization. In the cRPA method, the screening effect dominates the trend of $U_{\rm eff}$, whereas the stable $U^{\rm LR}_{\rm eff}$ indicates a balanced competition between the orbital localization and the screening effect.

\begin{figure}
\centering
\includegraphics[width=0.49\textwidth]{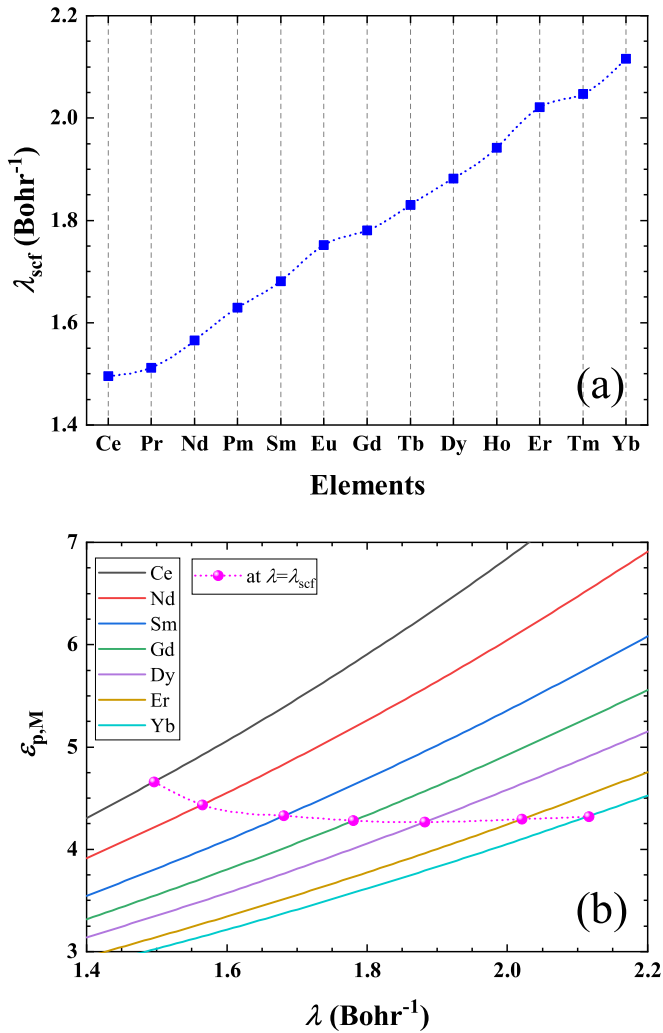}
\caption{(a) Screening parameter $\lambda_{\rm scf}$ (calculated with self-consistent charge density) of lanthanides in the LSCC approach. (b) Variation of macroscopic dielectric constant with respect to $\lambda$ (set manually) and elements. The magenta symbols represent $\lambda_{\rm scf}$.}\label{fig:lam}
\end{figure}

For LSCC, the screening parameter $\lambda$ depends on the charge density [see Eq. \eqref{TF-lam}]. Based on the Thomas–Fermi screening model, only the short-range metallic screening is considered in LSCC. Figure \ref{fig:lam}(a) shows that $\lambda_{\rm scf}$ gradually increases as the number of $f$-electrons increases, ranging from 1.50 to 2.12, where $\lambda_{\rm scf}$ denotes the screening parameter $\lambda$ calculated by self-consistent charge density. $\lambda$ can also be set manually. We report the macroscopic dielectric constant with respect to $\lambda$ for different elements (with different $f$-localized orbital) in Fig. \ref{fig:lam}(b). A larger $\lambda$ led to an enhanced short-range screening. However, the bare Coulomb interaction between correlated electrons is also enhanced at the same time. The macroscopic dielectric constant for LSCC is almost unchanged, and even slightly decreases, which is dominated by the correlated electrons.

\begin{figure}
\centering
\includegraphics[width=0.44\textwidth]{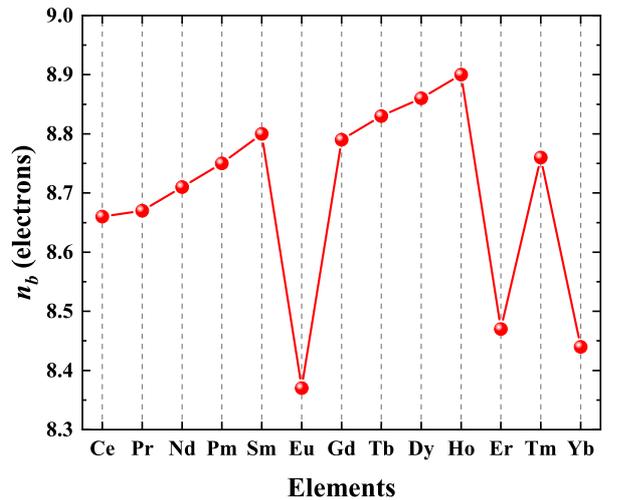}
\caption{Occupation number of bath electrons $n_{b} (= n_{\rm tot} - n_{f})$ of lanthanides.}\label{fig:occ}
\end{figure}

For LR, we calculated the occupation number of bath electrons (i.e., the result of total occupation number deducting the $f$-electrons, $ n_{b} = n_{\rm tot} - n_{f}$), as plotted in Fig. \ref{fig:occ}. With the increase of the radius of PAW augmented region (see Appendix), the occupation number of bath electrons gradually increases from Ce, but has three sharp drops in Eu, Er, and Yb. The trend of $n_b$ is same as $\varepsilon^{\rm LR}_{\rm p,M}$. LR approach is based on the perturbation-induced transfer of the occupation number, larger occupation number of bath electrons is likely to make charge transfer easier. However, the screening effect also depends on the correlated orbital, thus $\varepsilon^{\rm LR}_{\rm p,M}$ is not fully proportional to the occupation number of bath electrons $n_b$.
\begin{figure}
\centering
\includegraphics[width=0.44\textwidth]{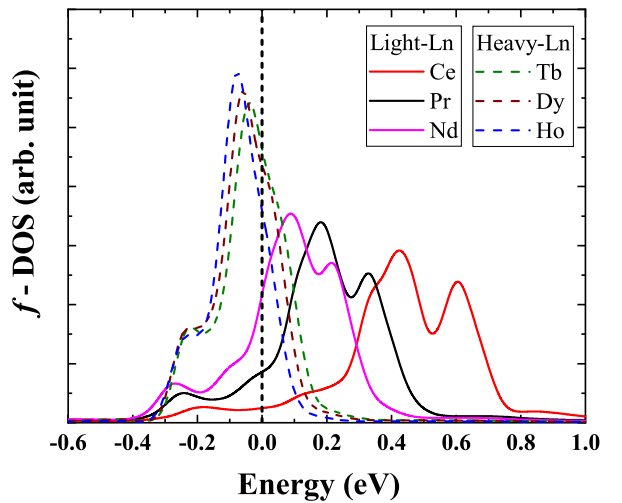}
\caption{Comparison of $f$ density of states. The solid lines represent $f$-DOS of typical light lanthanides: Ce, Pr, Nd. The chain lines represent $f$-DOS of typical heavy lanthanides: Tb, Dy, Ho. The black vertical line represents Fermi level.}\label{fig:dos1}
\end{figure}
For cRPA, the results are sensitive to the density of states, as cRPA is based on the transition between occupied–unoccupied electronic states (several eV), as seen in Eq. \eqref{screend-polar}. Nilsson \emph{et al.}\cite{Nilsson2013} noted that the density of states of $f$-electrons near the Fermi energy increases along the $4f$ series, leading to the enhancement of the polarization effect of $f$ and environmental electrons, and thus enhancing the screening. In our work, we also observed that the density of state of $f$-electrons near the Fermi energy increases with the filling of the $f$-shell. As an example, we showed typical cases in Fig. \ref{fig:dos1}. Meanwhile, under the uniform electron gas limit, the polarization is proportional to the charge density; thus, the occupation number of bath electrons, $n_b$, also affects the trend of $\varepsilon^{\rm cRPA}_{\rm p,M}$. We point out that the PBE exchange-correlation functional cannot qualitatively calculate the electronic state of the strongly correlated system, which leads to an excessive concentration of electronic states near the Fermi energy. This concentration of electronic states increases along the $4f$ series, which is one source of the excessive decrease of $\varepsilon^{\rm cRPA}_{\rm p,M}$. In the following section, we will simulate $U_{\rm eff}$ using the electronic state from PBE + $U$.

To the best of our knowledge, the judgement of different first-principles methods is still quite a complicated problem, and it is hard to say which scheme is best. However, we can make some suggestions for calculating the $U$ value of lanthanide metals based on our results and relevant simulations. From a theoretical point of view, LSCC is based on the Thomas–Fermi metallic screening model, and LR relies on the screening at the DFT level; both models are likely to provide a more appropriate description of metals. We note that $U$ values of 5.0–9.0 eV give results that are close to the experimental measurements.\cite{Harmon1995, Mohanta2010, Locht2016} In particular, the cohesive properties and spectra properties are in good agreement with the observations when an artificial $U$ value of 7.0 eV is set for all lanthanide metals in the DFT+DMFT simulation.\cite{Locht2016} The correlation strengths calculated by the LSCC and LR methods in this work fall within the 5.0–9.0 eV range. However, the simulation by cRPA probably needs a more sophisticated self-consistent scheme.\cite{Moree2018}

Lanthanide-based compounds are of great interest owing to their important roles in real applications. However, because of their complex electronic structure, the performance of correlation strength will be more complicated; therefore, they are not considered in this work. According to our results, cRPA is more sensitive to the screening effect causing by the chemical environment. Additionally, it has been reported that cRPA is suitable for gapped systems.\cite{vanLoon2021} Therefore, we believe that cRPA has great potential in the simulation of lanthanide-based compounds. We will work on correlation strength simulations targeted at compounds in the future.

\subsection{Sensitivity of $U_{\rm eff}$ to initial electronic state}
\label{Sensitive}
\begin{figure}
\centering
\includegraphics[width=0.44\textwidth]{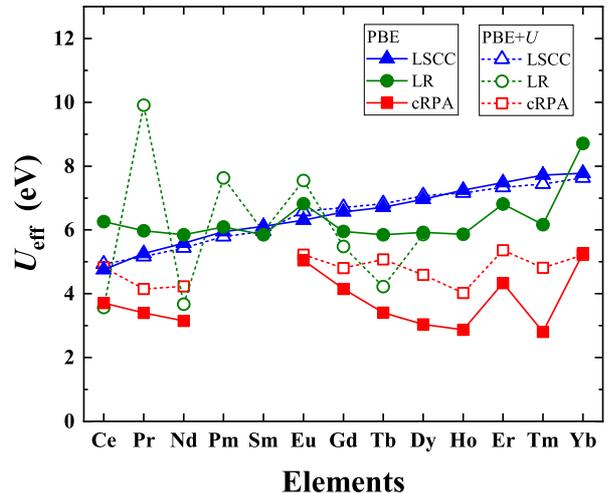}
\caption{Comparison of calculated $U_{\rm eff}$ value of lanthanides with initial electronic states from PBE and PBE+$U$ ($U$ = 6.0 eV). The filled symbols represent the results obtained with PBE electronic state, and open for the results obtained with PBE+$U$ electronic state.}\label{fig:sensi}
\end{figure}

The simulations above are all based on the same initial electronic states, which are obtained from PBE. As PBE (together with other LDA and GGA functionals) is often considered inappropriate for strongly correlated materials, it is necessary to determine how these $U$ simulation approaches perform from a more ``reasonable'' initial point. In Fig. \ref{fig:sensi}, the results calculated using PBE electronic states are compared with those calculated using PBE+$U$ ($U$ = 6.0 eV) electronic states. Using the PBE+$U$ electronic states, $U^{\rm cRPA}_{\rm eff}$ changes dramatically (compared with $U^{\rm cRPA}_{\rm eff}$ using PBE electronic states), whereas the LSCC approach only changes slightly (maximum change of 0.278 eV). It could be observed that $U^{\rm LR}_{\rm eff}$ also has large variation, except for Sm, but not monotonically.

\begin{figure}
\centering
\includegraphics[width=0.44\textwidth]{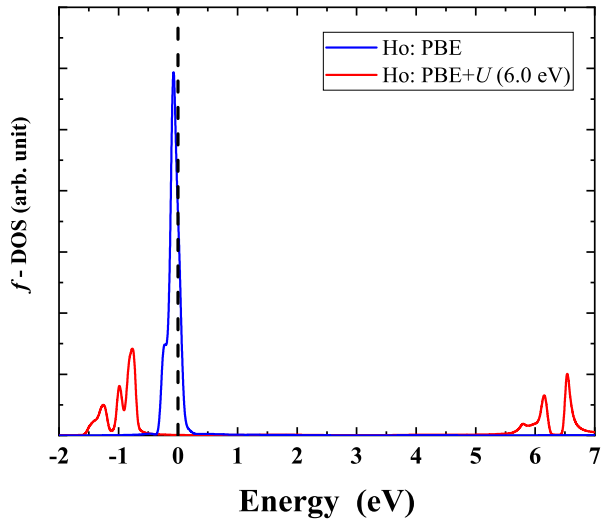}
\caption{Comparison of $f$-electron density of states. The blue line represents the $f$-DOS of Ho from PBE, and the red line represents the $f$-DOS of Ho from PBE+$U$ ($U$ = 6.0 eV). The black vertical line represents the Fermi level.}\label{fig:dosu}
\end{figure}

We speculate that the decrease in the density of states of $f$-electrons near the Fermi energy leads to the increase in $U^{\rm cRPA}_{\rm eff}$. Taking Ho as an example in Fig. \ref{fig:dosu}, we observe a significant decrease in the density of states of $f$-electrons near the Fermi energy when using PBE+$U$. For Eu and Yb, the electronic state obtained by PBE+$U$ is still large near the Fermi energy, whereas that obtained by PBE is not large near the Fermi energy. As the PBE+$U$ method mainly changes the occupation number and the density of states, the localized orbital is almost unchanged, and the behaviors of cRPA and LSCC are consistent with the characteristics we analyzed earlier.

We note that the process of LR self-consistent calculation is different from the cRPA method.\cite{Kulik2006} In the LR approach, the $U$ value is taken as the coefficient for the correction of the total energy with respect to the number of electrons, and the exact total energy varies linearly with respect to the number of electrons. In LDA/GGA, the number of electrons has a nonlinear contribution to the total energy, which is mainly a quadratic term brought by the Hartree term. 
After correction, the $U$ value obtained by LR calculation will be reduced accordingly. However, the $f$-electron orbitals are nearly degenerate in the standard $U=0$ PBE calculation, and perturbation is more likely to cause the variation of the occupations. In contrast, in the PBE+$U$ calculation, the corresponding energies of filled and unfilled $f$-electrons near the Fermi level become larger (e.g., see Fig. \ref{fig:dosu}), and the occupation is hard to change under the same perturbation.

Our results show remarkable variation between the $U_{\rm eff}$ from different initial electronic states, especially in cRPA, proving the necessity of understanding of the relation between DFT and the correction methods with $U$ (such as DFT+$U$). If methods like DFT+$U$ are treated as self-interaction-correction methods for PBE or other functionals, one should take $U_{\rm eff}$ from the initial electronic state, which is calculated by only the functional to be corrected. If DFT+$U$ is treated as functionals that should be solved in a fully self-consistent way, including $U$, the final $U_{\rm eff}$ is somehow calculated from DFT+$U$ initial states. These two kinds of interpretation could lead to very different results, as can be seen in this work. As more and more works have adopted a self-consistent cRPA scheme in the calculation of $U$ value, this issue becomes more crucial and deserves further investigation.


\section{Conclusion}

In summary, to research the factors that affect the effective Coulomb interaction strength in different first-principles approaches, we systematically calculated $U_{\rm eff}$ for lanthanide metals using the LSCC, LR, and cRPA methods, and we investigated their different performances. It could be observed that the trend of $U_{\rm eff}$ (as the $f$ shell is filled) has obvious differences between the three approaches. The $U^{\rm LSCC}_{\rm eff}$ value gradually increases, the $U^{\rm LR}_{\rm eff}$ value is almost unchanged at about 6.0 eV, and $U^{\rm cRPA}_{\rm eff}$ shows a two-stage decreasing trend in light lanthanides and heavy lanthanides, with an obvious change (1–2 eV) in lanthanides. Meanwhile, there are three abrupt jumps in Eu, Er, and Yb for both $U^{\rm LR}_{\rm eff}$ and $U^{\rm cRPA}_{\rm eff}$.

We analyzed these different trends based on the orbital localization and the screening effect. The rise of $f$-orbitals localization with increasing nuclear charge contributes to the increase of $U_{\rm eff}$, while the screening effect is also stronger in both light lanthanides and heavy lanthanides. The competition between the two factors is the main mechanism of the trends seen in the different approaches. We conclude that the trend of LSCC is dominated by the localization of $f$-orbitals, whereas that of cRPA is dominated by the screening effect, and $U^{\rm LR}_{\rm eff}$ is the balance of the two factors. Additionally, the sudden decrease of bath occupation number in Eu, Er, and Yb is related to the sharp changes of $U^{\rm LR}_{\rm eff}$ and $U^{\rm cRPA}_{\rm eff}$.

In addition, we simulated the dependence of $U_{\rm eff}$ on electronic state. The results show that $U^{\rm LR}_{\rm eff}$ and $U^{\rm cRPA}_{\rm eff}$ are sensitive to electronic state due to their stronger dependence on the screening effect. Thus, a self-consistent scheme for LR and cRPA is particularly important.

The behavior of different first-principles approaches for $U_{\rm eff}$ can be used to guide the choice of suitable $U_{\rm eff}$ parameter. A similar analysis can also be applied in other strongly correlated systems and other $U_{\rm eff}$ simulation approaches. A comparison of different first-principles approaches may provide a new perspective for the choice of Coulomb interaction strength in future research.

\begin{table}[tbp]
\caption{\label{tab:para} Values of experimental fcc primitive cell volumes and radius of PAW-augmented region for lanthanide metals.}
\begin{tabular*}{0.45\textwidth}{@{\extracolsep{\fill}} c|cc}
\hline\hline
Elements  &Volumes (\AA$^3$) &PAW radius (\AA)\\
 \hline
Ce        &34.34  &1.487  \\
Pr        &34.55  &1.492  \\
Nd        &34.02  &1.503  \\
Pm        &33.71  &1.513  \\
Sm        &33.29  &1.524  \\
Eu        &48.24  &1.529  \\
Gd        &33.19  &1.545  \\
Tb        &32.07  &1.550  \\
Dy        &31.56  &1.561  \\
Ho        &31.17  &1.572  \\
Er        &30.67  &1.386  \\
Tm        &30.09  &1.487  \\
Yb        &41.91  &1.498  \\
\hline\hline
\end{tabular*}
\end{table}

\section{Acknowledgement}

We thank Hong Jiang, Hua-Jie Chen, Xing-Yu Gao, Jie Sheng, Bo Sun, Ming-Feng Tian, and Jian-Zhou Zhao for helpful discussions. This work was supported by the National Nature Science Foundation of China (No. U1930401, No. 12004048, No. 12204033, and No. 11971066), the National Key Research and Development Program of China (No.2021YFB3501503), and the Foundation of LCP. We thank the Tianhe platforms at the National Supercomputer Center in Tianjin.

\section{Author Declaration}
\subsection*{Conflict of Interest}
The authors have no conflicts to disclose.
\subsection*{Author Contributions}
\textbf{Bei-Lei Liu}: Investigation (equal); Writing - original draft (equal); Writing - review \& editing (equal).
\textbf{Yue-Chao Wang}:
Supervision (equal); Conceptualization (equal); Investigation (equal); Writing - original draft (equal); Writing - review \& editing (equal).
\textbf{Yu Liu}:
Supervision (equal); Conceptualization (equal); Writing - original draft (equal); Writing - review \& editing (equal).
\textbf{Yuan-Ji Xu}:
Validation (equal); Writing - original draft (equal); Writing - review \& editing (equal).
\textbf{Xin Chen}:
Validation (equal); Writing - review \& editing (equal).
\textbf{Hong-Zhou Song}:
Validation (equal); Writing - review \& editing (equal).
\textbf{Yan Bi}:
Validation (equal); Writing - review \& editing (equal).
\textbf{Hai-Feng Liu}:
Validation (equal); Writing - review \& editing (equal).
\textbf{Hai-Feng Song}:
Validation (equal); Writing - review \& editing (equal).

\section{Data Availability}
The data that support the findings of this study are available from the corresponding author upon reasonable request.

\section{APPENDIX: VALUES OF VOLUME AND PAW PARAMETERS}
\label{app}
Details of the experimental fcc primitive cell volumes at ambient pressure and temperature\cite{McMahan2009}, as well as the radius of PAW-augmented regions for lanthanide metals, are presented in Table \ref{tab:para}.


\clearpage

\end{document}